\documentclass[12pt]{article}
\input epsf.sty
\begin{document}

%%Useful symbols%%%%%%%%%%%%%%%%%%%%%%%%%%%%%%%%%%
\def\CA{{\cal A}}
\def\CB{{\cal B}}
\def\CC{{\cal C}}
\def\CD{{\cal D}}
\def\CE{{\cal E}}
\def\CF{{\cal F}}
\def\CG{{\cal G}}
\def\CH{{\cal H}}
\def\CI{{\cal I}}
\def\CJ{{\cal J}}
\def\CK{{\cal K}}
\def\CL{{\cal L}}
\def\CM{{\cal M}}
\def\CN{{\cal N}}
\def\CO{{\cal O}}
\def\CP{{\cal P}}
\def\CQ{{\cal Q}}
\def\CR{{\cal R}}
\def\CS{{\cal S}}
\def\CT{{\cal T}}
\def\CU{{\cal U}}
\def\CV{{\cal V}}
\def\CW{{\cal W}}
\def\CX{{\cal X}}
\def\CY{{\cal Y}}
\def\CZ{{\cal Z}}

%macros
\newcommand{\todo}[1]{{\em \small {#1}}\marginpar{$\Longleftarrow$}}
\newcommand{\labell}[1]{\label{#1}\qquad_{#1}} %{\label{#1}} %
\newcommand{\ads}[1]{{\rm AdS}_{#1}}
\newcommand{\SL}[0]{{\rm SL}(2,\IR)}
\newcommand{\cosm}[0]{R}
\newcommand{\tL}[0]{\bar{L}}
\newcommand{\hdim}[0]{\bar{h}}
\newcommand{\bw}[0]{\bar{w}}
\newcommand{\bz}[0]{\bar{z}}
\newcommand{\be}{\begin{equation}}
\newcommand{\ee}{\end{equation}}
\newcommand{\lp}{\lambda_+}
\newcommand{\bx}{ {\bf x}}
\newcommand{\bk}{{\bf k}}
\newcommand{\bb}{{\bf b}}
\newcommand{\BB}{{\bf B}}
\newcommand{\tp}{\tilde{\phi}}
\hyphenation{Min-kow-ski}

%%Commonly used text expressions%%%%%%%%%%%%%%%%%%
\def\ie{{\it i.e.}}
\def\eg{{\it e.g.}}
\def\cf{{\it c.f.}}
\def\etal{{\it et.al.}}
\def\etc{{\it etc.}}

%%Commonly used constants and symbols%%%%%%%%%%%%%%%%%%%%%%%%%
\def\varep{\varepsilon}
\def\del{\nabla}
\def\grad{\nabla}
\def\tr{\hbox{tr}}
\def\perp{\bot}
\def\half{\frac{1}{2}}
\def\p{\partial}

%%%%%%%%%%%%%%%%%%%%%%%%%%%%%%%%%%%
% Erich's macros:
\renewcommand{\thepage}{\arabic{page}}
\setcounter{page}{1}

\rightline{}
\vskip 1cm
\centerline{\Large \bf Metabolically Efficient Codes in The Retina}
\vskip 1cm

\renewcommand{\thefootnote}{\fnsymbol{footnote}}
\centerline{{\bf Vijay
Balasubramanian${}^{1}$\footnote{Both authors contributed equally to 
this work.} and
Michael J Berry II${}^{2*}$,
}}
\vskip .5cm
\centerline{${}^1$\it David Rittenhouse Laboratories, University of
Pennsylvania}
\centerline{\it Philadelphia, PA 19104, U.S.A.}
\centerline{Email: {\it vijay@endive.hep.upenn.edu}}
\vskip .5cm
\centerline{${}^2$\it Department of Molecular Biology, Princeton 
University,
}
\centerline{\it Princeton, NJ 08544, U.S.A.}
\centerline{Email: {\it berry@princeton.edu}}

\vskip 0.15in 

\centerline{April 26, 2001}

%%%%%%%%%%%%%%%%%%%%%%%%%  Abstract (BEGIN)  %%%%%%%%

\begin{abstract}
We tested the hypothesis that the neural code of retinal ganglion cells is
optimized to transmit visual information at minimal metabolic cost.  Under
a broad ensemble of light patterns, ganglion cell spike trains consisted of
sparse, precise bursts of spikes.  These bursts were viewed as independent
neural symbols.  The noise in each burst was measured via repeated
presentation of the visual stimulus, and the energy cost was estimated from
the total charge flow in a biophysically realistic model of ganglion cell
spiking.  Given these costs and noise, the theory of efficient codes
predicts an optimal distribution of symbol usage.  Symbols that are either
noisy or metabolically costly are suppressed in this optimal code.  We
found excellent qualitative and quantitative agreement with the measured
distribution of burst sizes for ganglion cells in the tiger salamander
retina. \\

%\vspace{0.15in}

%\noindent{\bf {\small Keywords:}}
%\parbox[t]{5in}{{\it neural code; information theory; metabolic
%efficiency; \\
%retinal ganglion cell; noise; bursts}}
\end{abstract}

\setcounter{footnote}{0}
\renewcommand{\thefootnote}{\arabic{footnote}}

\newpage
%%%%%%%%%%%%%%%%%%%%%%%%%  Abstract (END)  %%%%%%%%

\leftline{{\bf {\large Introduction}}}
\vspace{0.15in}

Optimization principles remain a fundamental approach to understanding
the structure and function of the nervous system.  By quantifying the
performance of neurons, information theory provides very general
tools for constructing such principles.   An intuitive design, especially
for the sensory periphery, is that a neuron should maximize the mutual
information between input and output, subject to realistic constraints. 
Noise, either in the input or added by the neuron, is a fundamental
constraint.  Several researchers (\cite{barlow}; \cite{atick}; \cite{van})
have had success predicting how the receptive fields of retinal neurons
should be organized to maximize the transmission of information about
natural visual images.  This success might lead one to imagine that many
neurons strive to maximize information transmission.

However, there is a serious difficulty: spiking neurons, which
constitute the vast majority of neurons in the vertebrate brain,
produce action potentials at average rates far below the rate that
maximizes transmitted information.  A simple estimate can be made as
follows.  Since the maximum firing rate is set by the refractory
period $\mu \sim 5$ ms, an axon would achieve its highest entropy by
firing at a rate that gives a 1/2 probability of spiking in each
possible time bin, $r^* = 1/(2\mu) \sim 100$ Hz.  Timing jitter can be
expected to increase the effective bin size for spiking beyond $\mu$ and
hence lower the firing rate that achieves maximum information
transmission.  However, measurements from a number of neurons under
realistic stimulation show timing precision of $<$ 5 ms (\cite{bwm1};
\cite{rel1}; \cite{buracas}). Thus, one expects that most neurons would
achieve maximal information transmission at a firing rate close to $r^*
\sim 100$ Hz.  Instead, many neurons fire much less frequently: for
instance, ganglion cells in the salamander retina fire 1 to 5 spikes/sec
under varied stimulation~\cite{bwm1}, and pyramidal neurons in the rat
cortex fire at 1 to 4 spikes/sec (\cite{pyram1}; \cite{pyram2}), to name
just a few.  Therefore, one must conclude either that these neurons are
not well designed for information transmission or that other
considerations are important.

In fact, neurons are some of the most energy intensive cells in the
body \\ \cite{mitochon}, and spikes comprise a significant portion of their
energy usage (\cite{siesjo}; \cite{ames2}; \cite{brain2}; \cite{atlaugh}).
This suggests  that neural codes might be constrained by a requirement of
energy efficiency.   Previous work has developed a general theory of
energy-efficient codes  that predicts the optimal usage distribution of a
set of discrete, independent   neural symbols, each subject to an energy
cost and corrupted by noise (\cite{levy}; \cite{bkb}; \cite{gonzalo}).  We used this
prediction to make a quantitative  test of the hypothesis that the neural
code of retinal ganglion cells is energy efficient. Under a wide variety of
stimulus conditions, the output of ganglion cells is a sparse set of
discrete firing events~\cite{bwm1}. Each firing event is a tight burst that
we view as a composite neural symbol characterized by its number of spikes.
The noise of each firing event was measured by repeated presentations of
the same stimulus, and the energy cost was estimated by simulations of the
ionic current flow during spiking. The theoretically optimal distribution
of burst sizes was found to be in both qualitative and quantitative
agreement with experiment.

\vspace{0.15in}
\leftline{{\bf {\large MATERIALS AND METHODS}}}

\vskip0.15in
{\leftline {\bf Experimental Methods}}

Experiments were performed on isolated retinas from the larval tiger
salamander (Ambystoma tigrinum tigrinum).  The retina was perfused
with Ringer's medium (110 mM NaCl, 22 mM NaHCO$_3$, 2.5 mM KCl, 1.6 mM
MgCl$_2$, 1 mM CaCl$_2$, 10 mM D-glucose), which was oxygenated by
continuously bubbling a 95\%/5\% O$_2$/CO$_2$ mixture through it.  Action
potentials from retinal ganglion cells were recorded using a
multi-electrode array (see~\cite{bwm1} and~\cite{meister} for further
details) while the preparation was at room temperature.  This
temperature is realistic for tiger salamanders in their natural
habitat~\cite{roth}.  The retina was stimulated with illumination from a
computer monitor that was focussed onto the photoreceptor layer.  Stimuli
were drawn from an ensemble of spatially uniform random flicker, in which
intensity values for the entire screen were chosen every 30 ms from a
Gaussian distribution.  The mean light level was in the phototopic range
for salamander (7 mW/${\rm m}^2$), and the temporal root-mean-squared
fluctuation (contrast) was 35\% of the mean.  This level of contrast
approximates that found under more natural viewing
conditions~\cite{schaaf}.  Similar experiments were performed using
spatially modulated stimuli, in which square regions of the computer
monitor flickered independently, yielding ganglion cell responses similar
to the present study (data not shown).

Spike trains were recorded during 30 or 60 repeated presentations of
stimulus sequences lasting from 120 s to 540 s.  Total recording times
ranged from 120 min up to 270 min, with an average of 222 min.
Multiple tests of stationarity were employed to screen for ganglion
cells whose function remained constant during the experiment.  The
overall response, quantified by the average firing rate, remained
stable to better than 5\% (average stability over all ganglion cells,
3.6\%).  The noisiness of the response, quantified by the timing
precision and number precision (see~\cite{bwm1} for details), remained
constant to better than 10\% (averages 9.1\% and 8.2\%, respectively).
A total of 12 ganglion cells passed these tests of stationarity over
long enough recording times to be useful to our analysis.  They were
all fast OFF-type cells, identified by their unique reverse correlation
to random flicker stimulation ~\cite{smirnakis}.  These cells
resemble Y cells in the cat and M cells in the monkey.  In the tiger
salamander, as in most amphibians, the overwhelming majority (80\% -
90\%) of ganglion cells are OFF-type (\cite{smirnakis}; ~\cite{roth}).

\vskip0.15in
{\leftline {\bf Firing Event Identification}}

Under random flicker stimulation, salamander ganglion cells respond in
sparse, highly precise firing ``events" \cite{bwm1}. Most events were
easily identified as single peaks in the firing rate flanked by periods of
silence.  But in some instances, more ambiguous peaks in the firing rate
were observed.  In order to provide a consistent definition of firing
events, we identified event boundaries as reproducible minima in the
firing rate.  For each minimum $\nu$, the ratio of the firing rate at its
adjacent peaks ($p_1, p_2$) to that at the minimum was required to exceed
a threshold value, $\sqrt{p_1\, p_2} / \nu \, \phi$, with 95\% confidence.
The threshold was set to $\phi=2$; event boundaries were rather
insensitive to its value (see~\cite{bwm1} for addition details).  Once
event boundaries were drawn, all spikes between succesive boundaries were
assigned to a single event.  We denote the number of spikes at event
number $a$ on stimulus trial $b$ as $n_{ab}$.  If no spikes were observed
in a given trial, $n_{ab}$ was zero.  The average number of spikes in that
event is $N_a = \langle n_{ab} \rangle_b$, and the variance is ${\rm
var}(N_a) = \langle n^2 \rangle - \langle n \rangle^2$.

\vskip0.15in
{\leftline {\bf Energy Efficient Neural Codes}}

The properties of an optimally energy efficient neural code were
defined in earlier work (\cite{levy}; \cite{bkb}; \cite{gonzalo}).  Here, we 
imagine that a neural
circuit maps stimuli deterministically onto a discrete set of desired
neural symbols $\{Y\}$, and that noise is added before the output of actual
neural symbols $\{Z\}$ (see Fig.~\ref{modelnoise}A).  The goal of an energy
efficient code is to find the
probability $p_k$ of using each output symbol $z_k$ that optimizes
information transmission per unit metabolic cost. In the absence of
noise, the solution is:
\begin{equation}
p_k = e^{-\beta E_k}
\end{equation}
where $\beta$ is set such that $\sum_k e^{-\beta E_k} = 1$.  However,
noise profoundly alters the optimal symbol distribution~\cite{bkb} and
should therefore be carefully measured in a biological system and
included in the theory.

Noise is expressed as a transition matrix $Q_{k|j}$ that describes the
likelihood that the intended message or presymbol $y_j$ emerges from
the channel as $z_k$.   Given this ``noise matrix'', an iterative
algorithm generalizing classic work by Arimoto and Blahut 
(\cite{arimoto}; \cite{blahut}; \cite{blahutbook})
computes the symbol  distribution that optimizes information
transmission at a fixed average cost $E$.  This algorithm is 
summarized as follows:
\begin{enumerate}
\item Choose arbitrary nonzero probabilities for the presymbols $q_j^{(0)}$
\item For iteration $t=0,1,2,...$ repeat:
\begin{enumerate}
\item
Find the probability that observed output symbol $z_k$ arose
from the presymbol $y_j$:
$
   \hat{P}_{j|k}^{(t)} \leftarrow
      \frac{q_j^{(t)} Q_{k|j}}{\sum_j q_j^{(t)}Q_{k|j}}
$
\item  Update the optimal distribution of presymbols according to
\begin{equation}
             q_j^{(t+1)} \leftarrow \frac{ e^{ -\beta^{(t+1)} E_j - \hat{H}_j^{(t)}}}{
                           \sum e^{- \beta^{(t+1)} E_j -\hat{H}_j^{(t)}}}
\end{equation}
with $\beta^{(t+1)}$ chosen numerically so that $\sum_j q_j^{(t+1)} E_j =
E$.   Here we defined
$ \hat{H}^{(t)}_j \equiv - \sum_{k} Q_{k|j} \log \hat{P}^{(t)}_{j|k}$.
\item If $q_j^{(t+1)}$ close to $q_j^{(t)}$ stop.
\end{enumerate}
\end{enumerate}
We call the final, optimal probabilities of presymbols $q_{j}$ and 
define $\hat{P}_{j|k}$ and $\hat{H}_{j}$ in terms of $q_{j}$.   The 
optimal probability of output symbol $z_{k}$ is then given by 
$p_{k} = \sum_{j} Q_{k|j} \, q_{j}$.

Using this algorithm, we can always find the probability distribution
of coding symbols that optimizes the mutual information they transmit
about a sensory stimulus at a fixed average cost $E$.  In bits, this
maximized mutual information is given by $I = -\sum_j q_j (\ln q_{j} +
\hat{H}_{j})$, while the average energy used per symbol transmitted is $E =
\sum_j q_j E_j$.  Numerical optimization of $I/E$ then yields the code
that maximizes information transmission per unit metabolic cost.  The
necessary ingredients of this model are the energy cost of each output
symbol, $E_k$, and the noise in each presymbol, $Q_{k|j}$.

The energy efficient codes derived in this manner have three salient
features that are relevant for this study.  First, more energetic symbols
are suppressed in the optimal distribution.  Second, noisier symbols are
also suppressed.  Both these facts are illustrated in panels B and C of
Fig.~\ref{modelnoise}.   Finally, a surprising property of  metabolically
efficient codes is that  the distribution over symbols that maximizes
bits per energy is  actually invariant under  rescalings of the energy
costs $E_n \rightarrow \lambda \, E_n$ where $\lambda$ is some fixed
number  (see \cite{bkb} for a proof). This fact allows us to eliminate
one parameter in our model; the linear energy model $E_n = a_1 + a_2 \,
n$ can be rescaled by a factor of $1/a_{1}$ to $E_n = 1 + b \, n$, while
leaving the optimal distribution invariant.  We interpret the
resulting parameter $b=a_{2}/a_{1}$ as the relative cost of a spike
compared to the baseline cost of keeping the cell alive.

\vskip0.15in
{\leftline {\bf Classifying Presymbols}}

In order to provide a consistent and statistically significant
classification of firing events into presymbols, we used the following
algorithm:

(1) Compare the number of observations of the most common burst size $B_1$
to the number of observations of the next most common $B_2$.  If $B_1$
exceeds $B_2$ with 95\% confidence (assuming Poisson counting statistics),
then the event arises from an integer presymbol corresponding to the
most common burst size.  Most events (68\%) were classified as arising
from integer presymbols.

(2) If $B_1$ does not exceed $B_2$, compare $B_1$ to the observations of the
third most common burst size $B_3$.  If $B_1$ exceeds $B_3$ with 95\%
confidence, then the event is a half-integer presymbol corresponding to
the two most common burst sizes.  The vast majority of these events were
split between successive spike counts, such as 1 and 2 spike bursts,
accounting for 28\% of all events.  The remaining 0.2\% of all events had
bimodal burst size distributions, often between 0 and 2 spikes.  Because
the latter events were so rare, we simply treated them as integer
presymbols corresponding to the most common burst size.

(3) If $B_1$ does not exceed $B_3$,  check whether the three most
common burst sizes have successive values.  If so,  check
if the most common burst size is flanked by the next two most
common.  In this case,  assume that the event is  a noisier integer
event corresponding to the most common burst size (2\% of all events
fell in this category). Otherwise, assume that this is a  noisier
half-integer event characterized by the two most common burst sizes
(2\% of all events fell in this category).

(4) If the three most common burst sizes were not successive, we could
not clearly characterize the presymbol.  As only 0.05\% of all events
were in this final category, we again assumed that they were noisy
integer events corresponding to their most common burst size.

The Results section motivates the classification algorithm above
in terms of the measured characteristics of burst size distributions
for different events.

\vskip0.15in
{\leftline {\bf Burst Size Distributions and Noise Matrices}}

For each ganglion cell, we compiled a burst size distribution $P(z)$ from
the spike counts for individual events and trials, $n_{ab}$.  Increasingly
large bursts were increasingly rare, and their probability consequently
had large uncertainty due to random error. Therefore, in our analysis we
only included bursts up to a maximum burst size $z_{\rm max}$.  We denote
the number of observations of burst size $z$ in a single stimulus trial by
$N(z)$. The probability that a randomly selected burst in the trial has
size $z$ then is $P(z) = N(z) / \sum_{z} N(z)$, where we only include $z$
up to $z_{\rm max}$.  Our criterion for choosing $z_{\rm max}$ was that
$\sim 98\%$ of all bursts were included. Over our population of ganglion
cells, $z_{\rm max}$ ranged from 4 up to 6 spikes with an average of 4.6,
and the fraction of all bursts thus included ranged from 96.8\% to 99.6\%
with an average of 98.7\%.

We compiled the noise matrix $Q_{z|y}$ by estimating which presymbol
$y_a$ gave rise to each firing event (see Classifying Presymbols) and
adding the histogram of $n_{ab}$ for event $a$ over repeated trials
$b$ to the row of $Q$ corresponding to $y_a$.  We denote the number of
observations of burst size $z$ given presymbol $y$ as $L_{z|y}$, and
the number of observations of presymbol $y$ as $M_y$.  Each row of
the noise matrix was separately normalized by the number of observed
presymbols, $Q_{z|y} = L_{z|y} / (M_y \, R)$, where $R$ is the number
of stimulus repeats.  This normalization gives $\sum_z Q_{z|y} = 1$
as it should in order that $Q_{z|y}$ describe the probability that a
given presymbol $y$ produces the output $z$.  As discussed above,
the measured burst size distribution was truncated at a given $z_{\rm
max}$ for each cell, which can create unwanted ``edge effects''
in the theoretical analysis of the energy optimal burst size distribution.
In order to avoid these edge effects, we kept burst sizes greater than
$z_{\rm max}$ in the noise matrix.  However, after the theoretical
prediction of the optimal burst size distribution $P^*(z)$ was obtained,
we truncated it at $z = z_{\rm max}$ and re-normalized to unity,
$\sum_{z=1}^{z_{\rm max}} P^*(z) = 1$.   This allowed us to compare the
predictions of the theory properly with the measured distribution $P(z)$.

\vskip0.15in
{\leftline {\bf Error Analysis}}

For making quantitative comparisons between optimal and actual neural
codes, we must treat the uncertainty in our measurements properly.
Assigning error bars to the burst size distribution is not trivial,
because repeated presentations of the stimulus do not yield independent
measurements of the burst size distribution: these measurements are related
to each other by the noise matrix.  However, we can use the fact that
random flicker is completely uncorrelated from one stimulus frame to the
next. Therefore, the stimulus patterns that cause each type of firing
event occur independently and with a constant probability per unit
time. The number of presymbols of each type elicited by a stimulus sequence
of fixed length will vary with Poisson statistics:
\begin{equation}
  {\rm Var}[ M_y] = M_y.
\label{varm}
\end{equation}
For each presymbol, variations in the number of output spikes observed
on different stimulus trials are also independent and obey Poisson
statistics.  This implies that variations in $L_{z|y}$, the number of
observations of
$z$ given presymbol
$y$, are also Poisson:
\begin{equation}
{\rm Var}[ L_{z|y} ] = L_{z|y}.
\label{varl}
\end{equation}
We can use eqs.~(\ref{varm}) and (\ref{varl}) to find the error ${\rm
Var}(N(z))$.  By definition of the noise matrix, $N(z) = \sum_y
Q_{z|y} \, M_y = (1/R) \sum_y L_{z|y}$.  Therefore:
\begin{equation}
{\rm Var}[ N(z) ] = (1 /R) \sum_y {\rm Var}(L_{z|y}) = (1/R) \sum_y
L_{z|y} = N(z) \, .
\end{equation}
In order to find the error in the probability distribution over burst
sizes $P(z)$, we use standard error propagation:
\begin{eqnarray}
P(z) &=& {N(z) \over  \sum_{z'} N(z')} \\
{\rm Var}[ P(z)] &=&
\left( {1 \over \sum_{z'} N(z')} + {N(z) \over (\sum_{z'}
N(z'))^{2} }\right) \, {\rm Var}[N(z)] \\
{} &=&
\left( \sum_{z'}  N(z') + N(z) \over (\sum_{z'} N(z') )^{2} \right) \,
N(z) \nonumber
\end{eqnarray}
As we have discussed, a crucial input to the theory is the
experimentally measured noise in each ganglion cell.   Poisson
variation in the number of presymbols $y$ and the resulting
observations of outputs $z$ imply that the entries in the noise
matrix $Q_{z|y} = L_{z|y} / M_y R$ have variances
\begin{equation}
     {\rm Var}[Q_{z|y}] =  {2\,L_{z|y} \over M_y \, R} = 2 Q_{z|y}
\end{equation}
where we have once again used standard error propagation. Since we use a
numerical algorithm to derive the metabolically optimal code, it is not
possible to derive an analytic expression for the error bars in the
theoretical prediction. Furthermore, doing a complete numerical error
propagation is prohibitively time consuming.  Instead, we have estimated
the effects of the measurement errors on the theory by propagating the
most important uncertainties, namely those in the diagonal entries of the
noise matrix, through the algorithm that determines the optimal
metabolically efficient code.  The resulting errors are reported in
Fig.~\ref{fits} by including the maximum and minimum deviations in the
predicted probability for each symbol when the diagonal entries of the
noise matrix were perturbed by plus or minus one standard deviation.  In
the limit of small errors, this min-max error is similar to the result of
standard error propagation.  Further discussion appears in the Results
sections below.

\vskip0.15in
{\leftline {\bf Ion Channel Simulations}}
\newcommand{\sod}{{\rm Na}}
\newcommand{\calc}{{\rm Ca}}
\newcommand{\pot}{{\rm K}}
\newcommand{\ka}{{\rm A}}
\newcommand{\kca}{{\rm K,Ca}}

We studied the energy cost associated with electrical activity using a
single compartment model of tiger salamander ganglion cells due to
Fohlmeister and Miller (FM)~\cite{fohlmil}.  This model included five
membrane currents: voltage gated sodium ($I_\sod$), voltage gated calcium
($I_\calc$), delayed rectifier potassium ($I_\pot$), inactivating
potassium ($I_\ka$), and calcium-gated potassium current ($I_\kca$).   In
addition to these ionic currents the model includes an effective leak and
a capacitance leading to the equivalent circuit equation:
\begin{eqnarray}
I_{{\rm stim}} = & C_m \,{dV \over dt} + \bar{g}_\sod \, m^3 \, h \, (V
- V_\sod) +
\bar{g}_\calc \, c^3 \, (V - V_\calc) \nonumber \\
&
+ (\bar{g}_\pot \, n^4 + \bar{g}_\ka \, a^3 \, h_\ka + g_\kca )
\, (V - V_\pot) + \bar{g}_{{\rm leak}} \, (V - V_{{\rm leak}})
\label{ioneq}
\end{eqnarray}
This equation holds for every patch of cell membrane; so we
specify conductances in units of mS/cm${}^{2}$ and therefore describe
the current density through the ganglion cell membrane.
The cell capacitance ($C_m$), the ${\rm Na}^{+}$ and  ${\rm K}^{+}$
reversal potentials ($V_\sod, V_\pot$) and the conductances
$\bar{g}_\sod, \bar{g}_\calc, \bar{g}_\pot, \bar{g}_\ka$
were all given by constants specified in Table 1 of ~\cite{fohlmil}.
The leak reversal potential was chosen as $V_{{\rm leak}} = -65 \, {\rm
mV}$, so that after a long period with $I_{{\rm stim}} = 0$, the membrane
approaches a resting potential of $V = V_{{\rm leak}}$.   The leak
conductance was chosen to be $\bar{g}_{{\rm leak}} = 0.05 \,
{\rm mS}/{\rm cm}^{2}$.   (This is the value given in the text
of~\cite{fohlmil} and not in their Table 1.)  The calcium reversal
potential  was modelled dynamically by the
Nernst equation:
\begin{eqnarray}
V_{\calc} &=& {R \, T \over 2 \, F} \ln\left[{[\calc^{2+}]_e \over
[\calc^{2+}]_i(t) }\right] \\
{d[\calc^{2+}]_i \over dt} &=& -{3 \, I_{\calc} \over 2 \, F \, r} -
{([\calc^{2+}]_i - [\calc^{2+}]_{{\rm res}} ) \over \tau_{\calc} }
\end{eqnarray}
Here $R$ is the gas constant, $T$ is temperature, $F$
is the Faraday constant, $r$ is the radius of the cell,
$[\calc^{2+}]_{e}$, $[\calc^{2+}]_{i}$, and $[\calc^{2+}]_{{\rm res}}$ are
the external, intercellular, and residual intracellular calcium
concentrations respectively.  $\tau_{{\rm Ca}}$ is the time constant
for the removal of calcium from the intracellular space either by pumping
out of the cell or be sequestering in internal stores. These parameters
were extracted from the text of~\cite{fohlmil} and are specified in the
table  below.  The conductance of the calcium gated potassium channel
varies with calcium concentration as:
\begin{equation}
g_\kca = \bar{g}_\kca {([\calc^{2+}]_i / [\calc^{2+}]_{{\rm diss}} )^2  \over 1
+ ([\calc^{2+}]_i / [\calc^{2+}]_{{\rm diss}} )^2}
\end{equation}
where $[\calc^{2+}]_{{\rm diss}}$ is the calcium dissociation constant for
the calcium gated potassium channel. Finally, the state variables of the
voltage gated channels obey Hodgkin-Huxley equations:
\begin{equation}
{dx \over dt} = -(\alpha_x + \beta_x) \, x + \alpha_x
\end{equation}
where $x$ is any one of ($m, h, c, n, h_\ka$).   The parameters $\alpha_x$ and
$\beta_x$ are themselves dependent on the membrane potential $V$ as 
described in
Fohlmeister and Miller.   (Note that the parameter $E$ in Fig.~1
of~\cite{fohlmil} should be understood as the membrane potential
$V$.)  In summary, we used the following parameters in our simulations:
\begin{center}
     \begin{tabular}{|l|l|l|l|}
	\hline
	$\bar{g}_{\sod} = 50 \, {\rm mS/cm}{}^2$ & $\bar{g}_{\calc} = 
2.2 \, {\rm mS/cm}{}^2$ &
	               $\bar{g}_{\pot} = 12 \,  {\rm mS/cm}{}^2$ &
	$\bar{g}_{\ka} =  36 \, {\rm mS/cm}{}^2$ \\
	\hline
	$\bar{g}_{\kca} = 0.05 \, {\rm  mS/cm}{}^2$ & $ \bar{g}_{{\rm 
leak}} = 0.05 \, {\rm
	mS/cm}{}^{2}$
	  & $V_{\sod} = 35 \, mV$ & $V_{\pot} = -75 \, mV$ \\
	  	  	\hline
	$V_{{\rm leak}} = -65 \, mV$ & $C_{m} = 1\,\mu {\rm F/ cm}{}^{2}$ &
	$T = 22 \, {}^{\circ}C$ & $r= 0.0025 \, cm$ \\
	\hline
	$RT/F = 25.43 \, mV$ & $Fr = 241.22 \, cm$ & $\tau_{\calc} = 50 \,
	ms$ & $[\calc^{2+}]_{{\rm diss}} = 1 \, \mu M $\\
	\hline
	$[\calc^{2+}]_{{\rm res}} = 0.1 \, \mu M$ & $[\calc^{2+}]_{{\rm e}} =
	1800 \, \mu M$ & {} & {} \\
	\hline
     \end{tabular}
\end{center}
With these parameters fixed, we studied the energetics of spiking by
stimulating the model with brief current pulses following long resting
periods.  The charge flow in each ionic channel was obtained by
integrating the corresponding current as a function of time.  The ATP
consumption in ganglion cell soma required for transporting this charge
will be linear in the amount of charge, allowing us to determine how the
energy cost depends on the number of spikes and time intervals between
spikes in a burst. The results of these simulations are described
in Results.

\vspace{0.15in}
\leftline{{\bf {\large RESULTS}}}

\vskip0.15in
{\leftline {\bf Burst size distributions}}

Recordings were made from ganglion cells in the larval tiger salamander
retina while stimulated with the diverse temporal patterns of light found
in spatially uniform flicker (see Materials and Methods).  This paper
describes results from 12 ganglion cells of the fast OFF type.  Under
random flicker stimulation, these ganglion cells respond in sparse, highly
precise episodes of firing (\cite{bwm1}; \cite{inform4};\cite{meister2}).
These firing ``events" were tight bursts containing one to eight spikes
with inter-spike intervals of $5$ to $10$ ms.  Their first spike was
elicited with timing that typically jittered by $\sim 4$ ms from one
repeated stimulus presentation to the next.  The total number of spikes in
an event typically varied by $\sim 0.5$ spike.  In between events, the
spontaneous firing rate was strictly zero.  Events occurred at rates of 1
to 3 per second, with inter-event time intervals broadly distributed from
50 ms up to 3 s.  Because of their high precision, each firing event can
serve as an individual visual message of high fidelity. Because of the
long inter-event time intervals and the consequently weak correlations
between successive firing events, each event can serve as an independent
visual message.  This led us to consider each firing event as a composite
coding symbol, characterized by its time of occurrence and total number of
spikes.

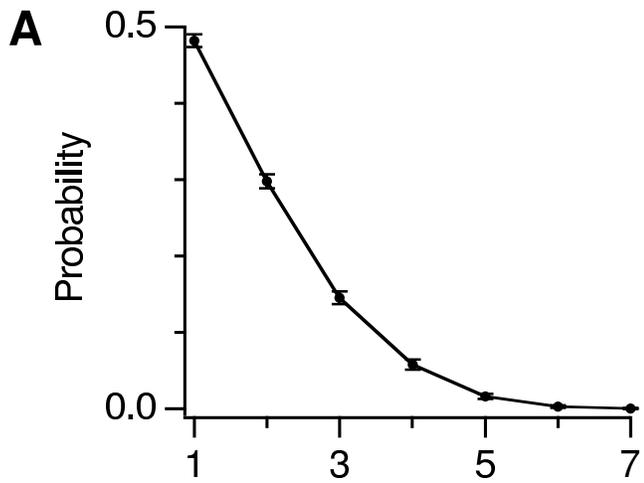
\begin{figure}[h]
%\vskip 0.5in
%\begin{center}
%\leavevmode
%\epsfxsize=6in
%\epsfysize=0.9in
%\epsffile{Fig1photo.eps}
%\end{center}
\caption{Burst size distributions for four representative salamander
ganglion cells, showing (A) very exponential shape, (B) significant
suppression of small bursts, (C) and (D) flattening of the distribution
for small bursts.}
\label{burstfig}
\end{figure}

Because events with different numbers of spikes can encode different
visual messages, this paper focuses on the distribution of burst sizes
chosen by retinal ganglion cells to represent a broad ensemble of stimuli.
Fig.~\ref{burstfig} shows such distributions for four cells along with
error bars that reflect the expected Poisson variation for stimuli drawn
from the same ensemble (see Materials and Methods). A characteristic
feature shared by all cells is the sharp falloff of bursts of large size.
An energy efficient code would behave in this way since large bursts are
expected to cost more energy.  Some of the cells have the surprising
feature of sharp suppression of small bursts (Fig.~\ref{burstfig}B), and
most show a flattening at small bursts (Figs.~\ref{burstfig}C
and~\ref{burstfig}D).  Naively, this seems to point away from energy
efficiency as an organizing principle for the retinal code since smaller
bursts cost less energy.  However, efficient coding also requires the
suppression of noisy symbols, and as we shall see, small bursts are
suppressed in precisely those cells where these bursts are not reliably
produced from trial to trial.

\vskip0.15in
{\leftline {\bf Structure of the model}}

In order to assess the energy efficiency of the retinal code, we developed
a simple model of a sensory system in which the properties of an
optimally energy efficient code could be solved ~\cite{bkb}.  In this
model, the action of a neural circuit was idealized as occurring in three
stages: stimuli in the external world $\{S\}$ are detected by an array of
sensors and converted into neural signals $\{X\}$, these signals are
encoded deterministically into a discrete set of composite symbols
$\{Y\}$, and finally noise is added before the output $\{Z\}$ emerges (see
Fig.~\ref{modelnoise}A).   For the retina, $S$ would be patterns of light
projected by the optics of the eye, $X$ would be the response of the array
of photoreceptors, and $Z$ would be the spike train produced by one or
more retinal ganglion cell.  The intermediate variable $Y$ is not directly
observable; we refer to its possible values $\{y_0,y_1,\cdots, y_n\}$ as
presymbols.

In the model, the output $\{Z\}$ is a string of discrete, independent
neural symbols $\{z_0,z_1,\cdots, z_n\}$.  Each symbol $z_k$ has an
energy cost $E_k$ associated with its production.  In the case of the
retina, individual firing events are viewed as the output symbols. 
Firing events are distinguished only by their number of spikes: $z_1$ is
a 1-spike burst, $z_2$ is a 2-spike burst, etc.  We add a symbol $z_0$
representing every 50 ms period of silence.   (Our results display
very little sensitivity to the value of the time bin used to discretize
periods of silence, as we have checked by picking a variety of time
bins.)  Because spiking represents a significant energy expenditure for
a neuron, firing events with different numbers of spikes have
significantly different energy costs. We will argue below that the
energy cost of a burst is linearly related to its number of spikes.

Given a set of costs $\{E_0, E_1, \cdots, E_n \}$ for the output symbols
$z_k$, and a noise matrix $Q_{k|j} = \Pr(z_k | y_j)$ which gives the
probability that a presymbol $y_j$ yields an output symbol $z_k$, we
can calculate the distribution of symbols $z_k$ which optimizes the
information transmitted per unit metabolic cost (see Methods)~\cite{bkb}.
The theory predicts that both noisy and energy expensive symbols are
suppressed in the optimal code. For example, Figs.~\ref{modelnoise}B  and
~\ref{modelnoise}C show a model with 6 input and 6 output symbols with
energies $E_k = k$.  In the absence of noise, the symbol $y_j$ is
transmitted as $z_k = y_j$, and bits per energy is maximized when the
usage  distribution of outputs $z_k$ is exponential ($p_k = e^{-\beta
E_k}$). Fig.~\ref{modelnoise}B shows that the addition of uniform noise
distorts the optimal symbol distribution, with a particularly strong
effect on the symbol $y_{1}$.  Fig.~\ref{modelnoise}C shows that adding
noise only to $y_{3}$ leads to marked suppression in its use, since this
symbol is then not a reliable conveyor of information.  Indeed, it is
shown in~\cite{bkb} that small changes in the noise can result in large
changes in the optimal distribution of output symbols.  So, to study
efficient coding by a ganglion cell, we must carefully characterize its
noise.

\vskip0.15in
{\leftline {\bf Noise in ganglion cells}}

Repeated presentations of the same input to the retina lead to a
distribution of output burst sizes for each firing event.
Fig.~\ref{defnoise}A plots the variance of the burst size distribution of
a given event against its mean. Most values have ${\rm
var}(N) < 0.5$, indicating that the number of spikes in a firing event is
precisely determined by the stimulus.  The solid line at the bottom of the
graph shows the theoretical lower bound on the variance of events with a
given mean, obtained when the event uses only outputs $n$ and $n-1$ with
the appropriate proportions. This theoretical lower bound drops to zero
for integer values of the mean spike number and rises to 0.25 for
half-integer values.  A corresponding pattern is seen for the actual
firing events: lower variance occurs near integer values of the mean spike
number.  This pattern is demonstrated by the thick line in
Fig.~\ref{defnoise}A, which is an average over the observed variance of
all events as a function of mean burst size.

This difference between high and low variance events arises
fundamentally from the nature of the mapping between the continuous
patterns of light occurring in the environment and the discrete outputs
of the retina. The set of all patterns of light constitutes a
high-dimensional space of stimuli, and we can view the retinal mapping as
dividing stimulus space into regions corresponding to presymbols \{Y\}
(see Fig.~\ref{defnoise}B for a schematic). Noise has the effect of
mapping a given stimulus onto a nearby location in stimulus space. When
the input starts out near the boundary between two presymbols, noise
will be much more likely to change the resulting output symbol than if the
stimulus is far from the boundary.  For example, Fig.~\ref{defnoise}B
shows regions of stimulus space corresponding to 1-spike and 2-spike
outputs.  Inputs that lie in the ``core'' (unshaded) of each region will
be mapped more reliably to output burst sizes, while inputs that lie in
the ``boundary'' (shaded) regions will be noisier.  As long as the outputs
of a neural circuit are discrete, there will exist boundary regions in
the input space that are more susceptible to noise.

If the retinal code is designed for efficient transmission of visual
information, its input-output map should attempt to reduce noise in
the output.  Fig.~\ref{defnoise}C displays the number of events for a
typical ganglion cell as a function of mean burst size.  Events with
an average integer number of spikes are more common than those with a
half-integer number.  As integer events tend to have lower noise, this
bias will have the effect of reducing the overall variability of
output symbols.  This dominance of ``core'' events over ``boundary''
events is not automatic.  For example, if the input-output map in
Fig.~\ref{defnoise}B contained a great many small disjoint regions
mapping to the same output, then the boundary regions could be greater
in size than the cores.  Furthermore, a simple model of the spiking
process, such as a linear filter followed by a static non-linear
function that is monotonic, would not emphasize integer events over
half-integer.  The fact that the stimulus-response mapping of ganglion
cells emphasizes core regions suggests that the retina can be
sensitive to minimizing its own noise.  It also suggests that we
should treat core presymbols distinctly from boundary presymbols when
characterizing the cell's noise.

\vskip0.15in
{\leftline {\bf Defining the noise matrix}}

The noise matrix in the model, $Q_{k|j}$, describes the transition
probability between presymbols $y_{j}$ and observed burst sizes
$z_{k}$. The presymbols $y_j$ are idealized, noise-free versions of the
circuit's desired output $z_k$.  As noise in the retina originates already
at the level of photon absorption by the photoreceptors, there is actually
no place in the retinal circuit where one can attempt to measure $y_j$.
However, the mapping from $X\rightarrow Y$ in Fig.~\ref{modelnoise} is
deterministic, so we can associate a unique value of $y_j$ with each
firing event.  Because every single observation of a firing event is
corrupted by noise, we must observe many instances of every firing event
to determine $y$. Our task is to use the distribution of burst sizes for
each firing event to infer that event's presymbol.

As the variability in spike number was very low, most firing events
had a narrow, unimodal distribution of burst sizes.  Some of these
distributions had their peak at a single spike count, while others had
a peak shared between two successive burst sizes.  The former case
corresponds to a core or integer presymbol, with a value equivalent to the
most common burst size (see Fig.~\ref{descnoise}A).  The latter
case corresponds to a boundary or half-integer presymbol, split between two
successive burst sizes (see Fig.~\ref{descnoise}B).  A small
percentage of events had bimodal burst size distributions (see
Fig.~\ref{descnoise}C) or had peaks spread among three or more burst
sizes.  These events were not well characterized by either integer or
half-integer presymbols, but they occurred so infrequently (less than
0.2\% of all events) that their awkward classification should not effect
our results appreciably. The algorithm used to classify presymbols is
described in Materials and Methods. Fig.~\ref{descnoise}D shows the
typical structure of a noise matrix.  The block on the ${\rm j}^{\rm th}$
row and ${\rm k}^{\rm th}$ column represents the value of $Q_{k|j}$ in
the shading depth -- darker colours represent  larger entries.  Notice
that integer presymbols are mostly mapped onto single output burst sizes,
while half-integer presymbols are mostly mapped onto two burst sizes.

\vskip0.15in
{\leftline {\bf Energetics of spiking}}

In addition to the noise matrix, the second ingredient we need in order
to compare the theory of energy-efficient codes to experimental burst-size
distributions is the energy cost of each neural symbol.  A significant
component of the total energy cost of a neuron results from its electrical
signaling (\cite{siesjo}; \cite{laughlin1}; \cite{atlaugh}).  In particular, 
action potentials are accompanied by large ionic currents that must then
be reversed by pumps at a cost in energy. For instance, ${\rm
Na}^{+}$/${\rm K}^{+}$ ATP-ase, the most common pump in the nervous
sytem, uses one molecule of ATP to pump 3 ${\rm Na}^{+}$ out of the cell
and 2 ${\rm K}^{+}$ in to the cell. By estimating the total charge flow
during a burst, we can consequently estimate the energy cost of that
burst.

We studied a single compartment model of the tiger salamander ganglion
cell developed by Fohlmeister and Miller (FM) (see Materials
and Methods).  Their model used six ionic currents with conductance
parameters set to closely match the cell membrane voltage measured during
spikes and bursts~\cite{fohlmil}.  We excited the FM model with
current pulses of increasing duration and amplitude to produce bursts of
action potentials. Fig.~\ref{spikecost}A shows an example of three
different current pulses and the resulting bursts of voltage spikes.
Fig.~\ref{spikecost}B shows the ionic current for each channel type.  The
total charge flow in each of these channel types was computed for a variety
of burst sizes by integrating the ionic current. The results are
displayed in Fig.~\ref{chargeflow}, along with a linear fit to the data
from the delayed-rectified ${\rm K}^{+}$ current.

Clearly the charge flow, and hence the ATP consumption, is linear with
burst size in the FM model and insensitive to the time intervals between
spikes in a burst.  What is more, the total charge flow associated with
action potentials greatly exceeds the resting charge leakage through the
cell membrane.  These results held for all the amplitudes and durations of
injected current that were studied.  While there are certainly other costs
and constraints relevant to biological information processing (see the
Discussion), this shows that the energy cost of producing bursts in real
ganglion cells is likely to be both large and linear in burst size.
Therefore, it is reasonable to assume such a linear energy cost model when
assessing the energy efficiency of the retinal code.

\vskip0.15in
{\leftline {\bf Theory meets experiment}}

Accordingly, we assume that a burst of size $n$ has a cost linear in
$n$: $E_n = E_0 + E_s \, n$.  Here, $E_0$ is interpreted as the cost of
50 ms of silence, and $E_s$ is the added cost of a single spike.  Since
codes that maximize bits per energy are invariant under rescalings of the
cost model: $E_n \rightarrow \lambda \, E_n$~\cite{bkb}, we can rescale the
costs by $1/E_0$ and arrive at
\begin{equation}
E_n = 1 + b \, n
\end{equation}
as a one-parameter model of the cost of a burst of size $n$.
The energy slope $b = E_s/E_0$ is the ratio of the cost of
producing a spike to the cost of 50 ms of silence.  Given  the
typical $\sim$5 ms refractory period between action potentials, a slope
of $b=1$ would imply that an action potential is of order 10 times as
expensive as an equal duration of silence.

With the measured noise and the above linear cost, we used the
algorithm presented in Materials and Methods to compute the
energy-efficient code for different slope parameters $b$.
The number of zero symbols $N(z_0)$ depends on the time bin used to
discretize silence.  Because our choice of a 50 ms time bin is arbitrary,
we left the zero symbol out of the burst size distribution that we
attempted to fit with our theory.  Furthermore, large bursts were too
infrequent to use in our analysis, so we truncated the distribution at a
largest burst size $z_{max}$ (see Materials and Methods). Therefore, after
computing the optimal distribution over all output symbols in our model
($0,1,2,\cdots$), we dropped the $0$ as  well bursts with sizes greater
than $z_{{\rm max}}$.     We then  renormalized the theoretical optimal
distribution over finite sized bursts ($1,2,\cdots , z_{{\rm max}}$). 
This enabled direct comparison with the experimentally measured
distribution of burst sizes.  For each cell in our data set we performed
fits of the theory to the experimentally measured distribution by varying
the energy slope $b$ and minimizing $\chi^2$. Excellent agreement was
found between theory and experiment as summarized in Fig.~\ref{fits}.

{\it Qualitative Agreements: } Figs.~\ref{fits} shows representative fits
for four cells and display excellent qualitative agreement.  For all
cells, the sharp falloff of large bursts was accurately reproduced by the
theory.  Most cells had some suppression of 1-spike bursts. 
Interestingly, these cells also had higher noise in the 1-spike bursts,
which were consequently suppressed in the theoretical efficient code.
Some cells showed a marked dip in the use of small bursts (e.g.
Fig.~\ref{fits}B).  As shown, such dips were also reproduced by the
theory, again because of the suppressed use of noisy symbols in an
efficient code.  It is striking that the decreased use of noisy symbols
that would be predicted by information theory is reproduced in the
experimental data. 

{\it Quantitative Agreements: } Fig.~\ref{fitparams}A shows the $\chi^2$
per degree of freedom plotted against the energy slope $b$ values obtained
from fits to each cell.    ($\chi^2$ per degree of freedom is
defined as $\chi^2/(N-1)$, where the total $\chi^2$ for the fit is
normalized by the number of fit points minus the number of parameters.
Whenever we refer to $\chi^2$ in the text it should be understood as
$\chi^2$ per degree of freedom.)  The open squares show the values
obtained by fitting with the empirically measured noise matrix.  Low
$\chi^2$ values are obtained for all cells.  The higher values
typically occur when the fit fails to match at one point leading to a
large $\chi^2$ contribution.  For instance, notice in Fig.~\ref{fits}C
that $\chi^2 = 7.2$ due to a large mistmatch in the 5-spike burst, even
though the fit of theory to experiment looks quite good to the eye.
 
Because each element of the experimentally measured noise matrix has error
bars, the theoretical burst size distribution also has some uncertainty. 
This uncertainty is shown as the range bars in Fig.~\ref{fitparams}A and
was computed by propagating the errors in the noise matrix numerically
through the theory (see Materials and Methods).  The large ranges again
occur because some perturbations of the noise matrix led to a bad fit at
one point. The ranges in the optimal symbol probabilities produced by
perturbing the noise matrix are displayed in Figs.~\ref{fits} as the
thinner upper and lower lines, showing that the fits remain very similar
under perturbations in the noise matrix.

The large variation in energy slope $b$ found within a population of
ganglion cells of nominally the same functional type may seem
surprising.  However, this group of cells also had a large range of
overall firing rates (roughly a factor of 10), despite the fact that the
stimuli were statistically identical for all of them.  If ganglion cells
are implementing an energy efficient code, one expects that a cell having
a small incremental cost of spiking (i.e., a small value of $b$) will
choose to fire often.  Similarly, large energy costs for spiking would
imply lower firing rates.  In fact, Fig.~\ref{fitparams}B shows that the
overall firing rate of a cell is inversely related to its energy slope
$b$, consistent with the principle of energy efficiency.

The good quantitative agreement between an optimally energy efficient code
and the actual retinal code requires that both the energy costs and noise
of neural symbols be taken into account.  To demonstrate this point, we
compared our model to two others: one which neglects the noise of
transmission through a channel, and another which includes the empirical
noise, but assumes all neural symbols have the same energy cost.  As shown
in Fig.~\ref{fitcompare}A, models neglecting the cost of transmission
(dashed line) fail to capture the rarity of large bursts, while models
neglecting noise (dotted line) miss the suppression of noisy small
bursts.  By contrast, the full model (solid line) captures both the
suppression of large, expensive bursts and small noisy ones as seen in
the data (dots with error bars). Fig.~\ref{fitcompare}B compares
$\chi^{2}$ of the fits to the data using the full and partial models. The
dark and open circles represent models  that neglect noise and energy
costs, respectively. It is clear that a model incorporating both factors
is a significant quantitative improvement.

\vskip0.15in
\leftline{{\bf {\large DISCUSSION}}}

In summary, we have argued that one aspect of the neural code of retinal 
ganglion cells, the distribution of burst sizes used to represent a broad
stimulus ensemble, is surprisingly sensitive to both the noise and the
energy cost of bursts.  Retinal ganglion cells suppress the bursts with
greater noise and greater metabolic cost, in both cases consistently with
the principle of maximizing visual information per unit metabolic cost. The
retina even appears to map stimuli preferentially onto less noisy firing
events (see Fig.~\ref{defnoise}C). By measuring the noise in each firing
event and estimating the energy cost up to one free parameter, we compared
the actual burst size distributions of ganglion cells to the distribution
that optimizes visual information per unit energy.  Excellent agreement was
found.  These results suggest that the retina strives to use an energy
efficient neural code.

Why should retinal energy efficiency matter to an animal, considering
that the retina comprises only a small fraction of most vertebrates'
bodies?  In fact, neurons are some of the most energy-intensive cells in
the body.  For instance, the human brain accounts for 20\% of daily
energy intake but only 2\% of body mass \cite{humanbrain}. Electrical and
chemical signaling has been estimated to account for up to 85\% of a
neuron's energy budget \cite{atlaugh}.  These data suggest that retinal
neurons should attempt to conserve energy.  Furthermore, the energy saved
in the retina by an optimal code may underestimate the total savings for
the animal.  Retinal spike trains lead to cascades of synaptic currents
and action potentials in the central brain, the costs of which might be
reduced by using an efficient retinal code.  Also, the retina employs much
of the same molecular machinery -- from ion channels to biochemical
signaling cascades -- found throughout the brain.  As a result, mutations
in this machinery that allow for greater energy efficiency in the retina
may simultaneously enable other parts of the brain to save energy.

Despite these reasons, one might wonder why the efficient codes studied in
this paper are important: photoreceptors are known to be metabolically
expensive, and optimal coding does not affect this initial stage of the
visual pathway.  Ames \cite{ames} has made a detailed analysis of energy
costs in the rabbit retina.  He finds that while the photoreceptors do
cost more ATP to run, the ganglion cell layer uses anaerobic respiration,
so that its cost in terms of glucose consumption is comparable to the
photoreceptor layer.  Since the ganglion cell layer does not receive a
direct blood supply in either the rabbit or salamander retinas, it is
likely that the ganglion cell layer of salamander uses anaerobic
respiration, too.  In addition, a simple estimate of the total membrane
area of the optic nerve in the salamander indicates that its energy cost
may exceed the entire retina for reasonable firing rates, and this cost
will be reduced by an efficient code.

The ability of neurons to use a metabolically efficient code must be
selected for by evolution.  We imagine that mutations which allow the
retina to use an energy efficient code would leave the animal with
identical visual behavior, but at a reduced metabolic cost.  Determining
the magnitude of the fitness advantage that arises from such a reduction in
metabolic cost is very difficult.  But even a slight fitness advantage can
come to dominate a population: if the fractional fitness advantage is $s$,
then the time necessary to spread through the population is $1/s$
generations \cite{popgen}.  Thus, a fitness advantage of one part in a
million requires only a few million years to become widespread, which is
fairly quick in evolutionary time.  Evidence that eyes can be a
significant expense to maintain is found from species of cave fish and
salamanders that have adapted to environments without light
(\cite{nelson}; \cite{roth}).  In these species, the eyes have been fully
or partially lost.  Some, like {\it Astyanax fasciatus mexicanus}, have
very close relatives that live outside caves and retain their eyes.

Our estimate of the energy cost of a burst left one free parameter, the
energy slope $b$, measuring the relative cost of spiking to silence.  This
parameter took on a wide range of values when the model was fit to our data
(see Fig.~\ref{fitparams}).  What is a reasonable value of $b$?  Our ion
channel simulations suggest that the ion flow during a spike is $\sim 100$
times the leak current during 50 ms.  However, the ganglion cell will have
baseline metabolic costs in addition to the leak current.  If we assume,
following Siesjo and Ames, that half of the energy cost for ganglion cells
is needed for baseline metabolism (\cite{siesjo}; \cite{ames2}), and the
other half for firing spikes at the observed rates of about 5 spikes/sec,
then we estimate $b \sim 4$.  However, we expect large variation from cell
to cell due to differences in dendritic morphology, axon diameter, and ion
channel densities \cite{roth}.  The large range of observed firing rates
(see Fig.~\ref{fitparams}A) suggests significant heterogeneity in our
population of ganglion cells.  In addition, one might interpret the
energy cost parameter as including expenses of other neurons, like
presynaptic bipolar cells or postsynaptic tectal cells, or as effectively
including other constraints.  It is worth considering whether these costs
are relevant to the retinal code, and how they should be folded into  our
analysis.  Clearly, more experimental data on the energy cost of neural
signaling is needed.

Our analysis would benefit from several extensions.   The first is
developing a model that explicitly represents time.  Because our model
treats the spike train merely as an ordered list of firing events, it leaves
out all of the information contained in the exact time at which each event
occurs.  In addition, we were forced to use an arbitrary time bin, 50 ms,
to identify periods of silence as a zero-spike burst, which prevented us
from predicting the overall firing rate of a ganglion cell.  Other
important extensions include testing energy efficiency for different 
functional types of ganglion cells, different species, and different
stimulus conditions, to assess the generality of our result.  It is of
particular interest to see whether a given ganglion cell can maintain an
energy efficient code when the statistics of its visual input change
significantly.

This article raises interesting questions.  We have shown that retinal
ganglion cells use a code that suppresses noisy neural symbols.  As the
statistics of the visual input (such as light level or contrast) change,
the noise also changes.  In order to remain efficient, a ganglion cell
would have to detect changes in its own noise and adjust its code
accordingly.  How could this happen? The only likely method is for a cell
to compare its responses to neighbors that are performing a similar
function. This would suggest that a certain amount of redundancy is
essential in ganglion cell function in order to enable adaptation to
noise.  If the picture of neurons as precise and robust
information-bearing channels is correct, we must understand the
mechanisms by which neurons detect and adapt to their own errors. 
Similarly, in the context of this paper, it is interesting to identify
neural mechanisms for adapting to changing metabolic conditions.

\vskip0.2in
\centerline{\bf Acknowledgments}
\medskip

We have enjoyed discussions with Larry Abbott, Bill Bialek, Kwabena
Boahen, E.J. Chichlinsky, Mike deWeese, Gonzalo Garcia de Polavieja,
Christof Koch, Simon Laughlin, Elad Schneidman, Peter Sterling, Chuck
Stevens, and Gerard Toulouse.  We are grateful to Don Kimber and
Markus Meister for collaboration on previous work directly related to
this article.  During this project {\small V.B.} was supported
initially by the Society of Fellows and the Milton Fund of Harvard
University and later by DOE grant DOE-FG02-95ER40893.  {\small M.B.}
was supported by the Pew Charitable Trust and the E. Mathilda Ziegler
Foundation.  {\small V.B.} is grateful to the ITP, Santa Barbara and
the Aspen Center for Physics for hospitality while this work was in
progress.

\vskip0.2in

%\newpage

%%Bibliography

\vfill

\newpage
\begin{center}
{\bf CAPTIONS}
\end{center}

\vskip0.5in

%\newpage

\vskip0.5in

%\newpage
\begin{figure}[h]
%\vskip 0.5in
%\begin{center}
%\leavevmode
%\epsfxsize=6in
%\epsfysize=0.9in
%\epsffile{Fig2photo.eps}
%\end{center}
\caption{(A) Simplified schematic of our model of a sensory system.
(B) and (C) Effects of noise in the model: the optimal usage probabilty of
each symbol is plotted for different levels of noise.
In both (B) and (C), there are 6 output symbols, and the energy cost of
symbol $k$ is $E_k = k$.  In (B), the noise matrix has the following
``diagonal'' form: $\Pr(z|y) = 0$, except for $\Pr(z=k|y=k) = 1 -
2p$ for $k=1\cdots6$; $\Pr(z=k+1|y=k) = \Pr(z=k-1|y=k) = p$ for $k=2\cdots
5$; $\Pr(z=2|y=1) = \Pr(z=5|y=6) = 2p$.  The optimal
usage distribution is exponential when $p=0$, showing suppression of
energetically expensive symbols.  As the noise increases, there is a
marked deviation from an exponential symbol distribution.  In (C), the
noise matrix is of the same form as in (B), but with $p=0.1$ for all $k$
except $k=3$, for which the noise probabilities are varied.  This panel
shows that noise in a single symbol strongly supresses the use of that
symbol in the optimal distribution.}
\label{modelnoise}
\end{figure}

%\newpage
\begin{figure}[h]
%\vspace*{2in}
%  \begin{center}
%   \leavevmode
% \epsfxsize=6in
% \epsfysize=8in
%   \mbox{\epsfbox{Fig3photo.eps}}
%\end{center}
\caption{(A) Variance of the spike count in a firing event
plotted as a function of average spike number for 1043 events. (B)
Deterministic mapping of stimulus to presymbol: noisy boundary versus
reliable core.  Visual stimuli occupy a large space, where each point
is a pattern of light intensity as a function of time (shown on far
left).  This stimulus space $\{ X \}$ is mapped deterministically into
presymbols $\{ Y \}$, which are in turn mapped  into the output symbols
$\{ Z \}$ by the noise matrix.   The figure shows two regions of
presymbol space which should be mapped into one or two spikes,
respectively, in the  absence of noise.   In the presence of noise,
stimuli that are mapped into the core of these  regions will be reliably
encoded into the same number of spikes, while stimuli  mapped into the
boundary regions as marked will be readily corrupted,  being transmitted
sometimes as  one spike and sometimes as two.    See the text for
discussion.  (C) Histogram of different presymbols used by one ganglion
cell.}
\label{defnoise}
\end{figure}

%\newpage
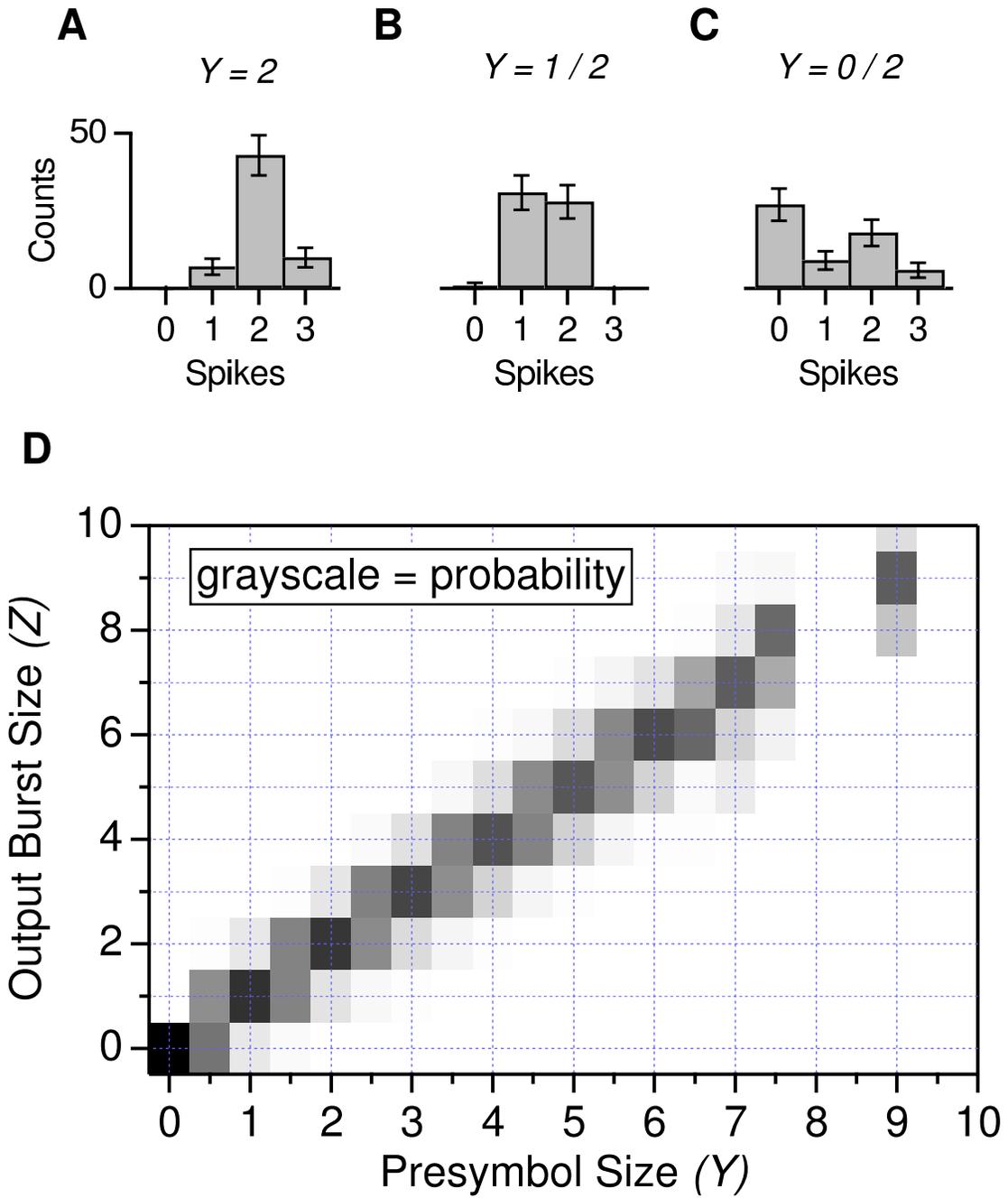
\begin{figure}[h]
%\vspace*{2in}
%  \begin{center}
%   \leavevmode
%   \epsfxsize=6in
%   \epsfysize=7.5in
%   \mbox{\epsfbox{Fig4Canvas.eps}}
%\end{center}
\caption{(A)--(C) Distribution of spike counts for three
representative firing events of types 2, 1/2, and 0/2. (D)
Depiction of the noise matrix for a ganglion cell with the
transition probability $Q_{z|y}$ for a presymbol $y$ to emerge as an
output symbol $z$ plotted in greyscale.}
\label{descnoise}
\end{figure}

%\newpage
\begin{figure}[h]
%\vskip 0.5in
%\begin{center}
%\leavevmode
%\epsfxsize=6in
%\epsfysize=7.5in
%\epsffile{Fig6photo.eps}              
%\end{center}
\caption{Ion channel simulations:  (A)  Regular spiking behavior
seen in the voltage response $V$ of the  Fohlmeister-Miller
model (top panel) to current pulses $I_{{\rm ext}}$ of varying durations
(bottom panel). (B) Associated currents through the cell membrane,
including voltage-gated sodium $I_{{\rm Na}}$, delayed rectifier
potassium $I_{{\rm K}}$, potassium A-current $I_{{\rm A}}$, voltage-gated
calcium $I_{{\rm Ca}}$, calcium-gated potassium $I_{{\rm K(Ca)}}$, and
leak $I_{{\rm leak}}$.}
\label{spikecost}
\end{figure}

%\newpage
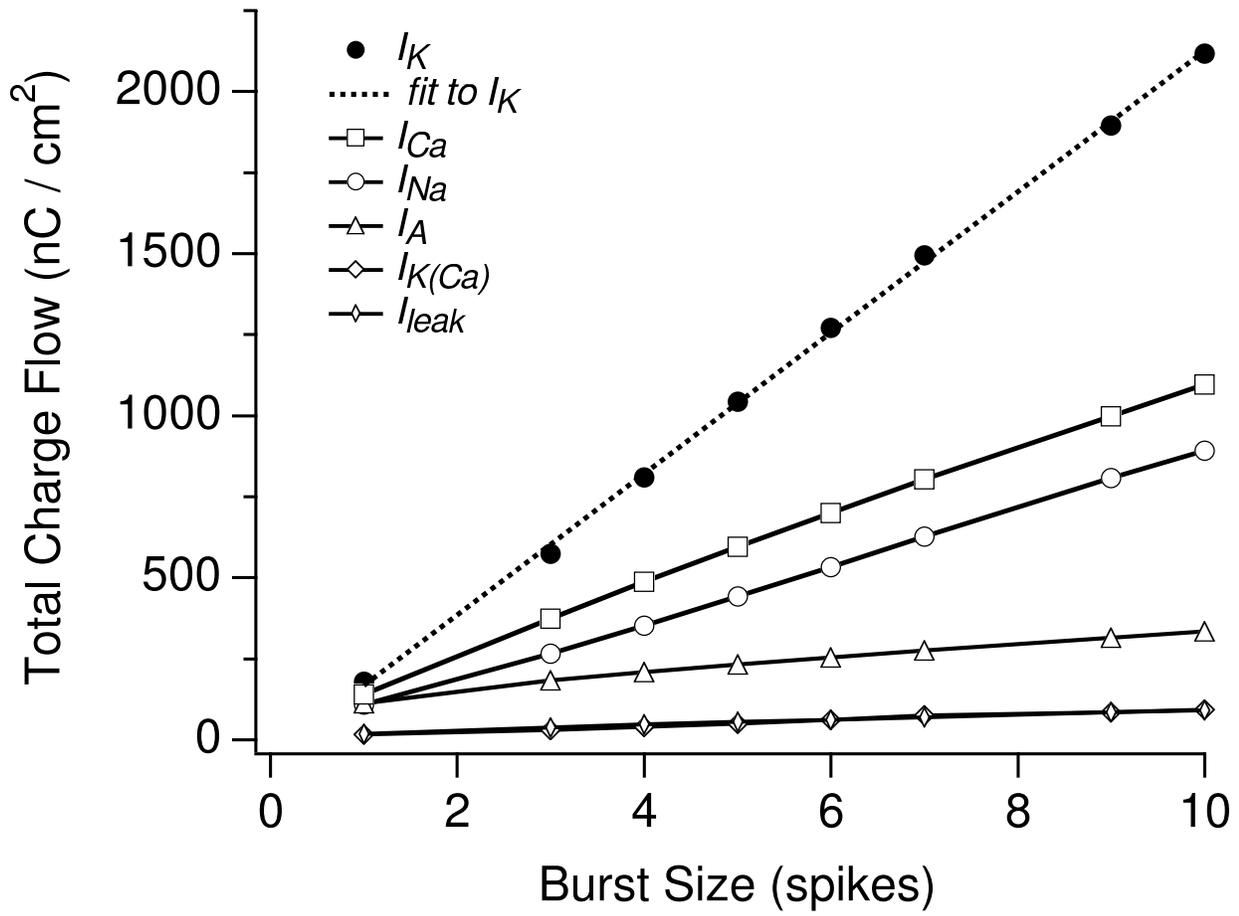
\begin{figure}[h]
%\vskip 0.5in
%\begin{center}
%\leavevmode
%\epsfxsize=6in
%\epsfysize=8in
%\epsffile{Fig7photo.eps}
%\end{center}
\caption{Total charge flow in the Fohlmeister-Miller model plotted as a
function of the number of spikes in a 50 ms period.  For the delayed
rectifier potassium current, a linear curve fit is shown as a dashed 
line.}
\label{chargeflow}
\end{figure}

%\newpage
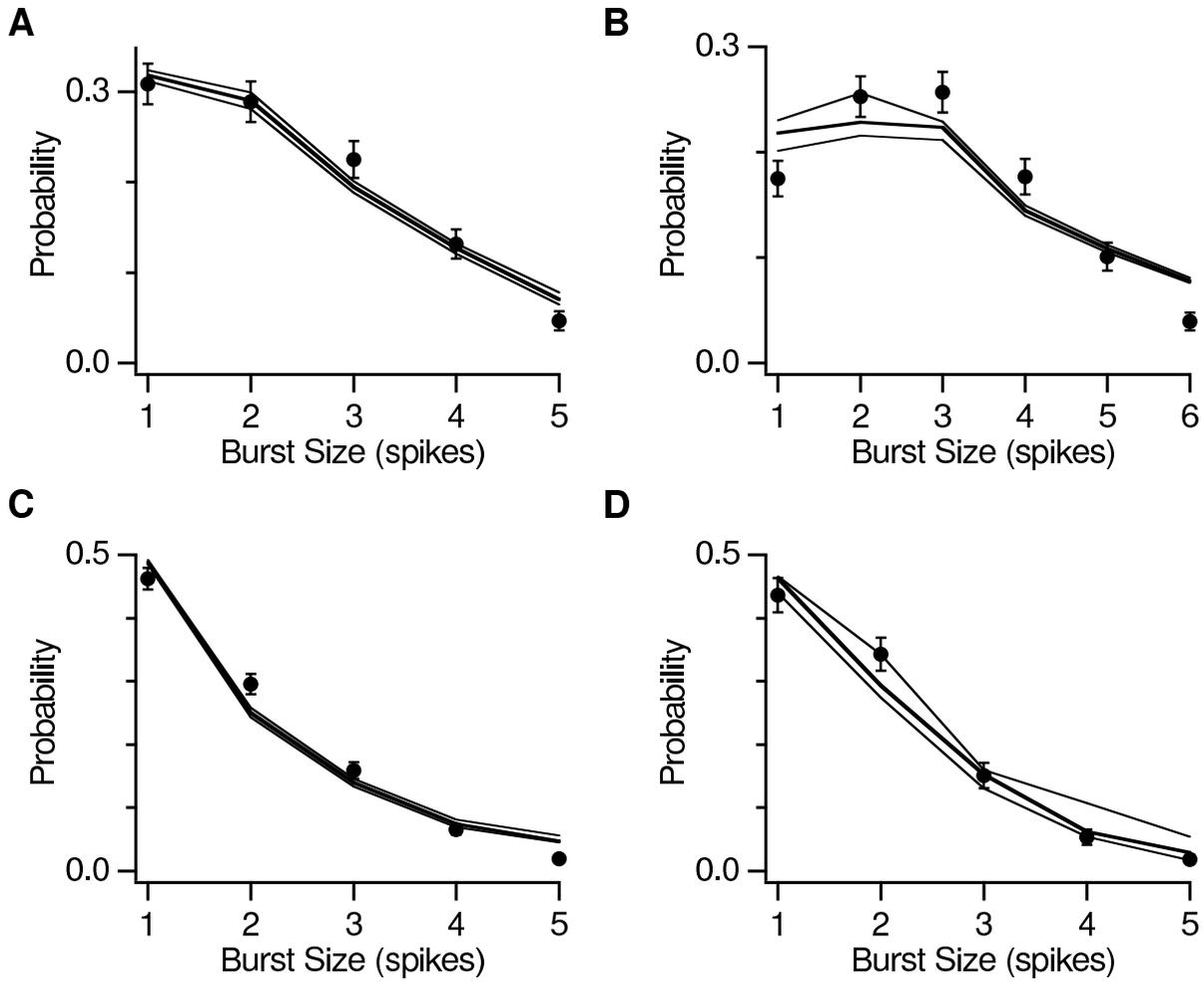
\begin{figure}[h]
%\vskip 0.5in
%\begin{center}
%\leavevmode
%\epsfxsize=6in
%\epsfysize=7.0in
%\epsffile{Fig5aphoto.eps}
%\end{center}
\caption{ Curve fits of an optimally energy efficient code to
the experimental burst size distribution for four cells made by finding
the best value of the energy cost parameter $b$.  Dots with error bars are
the experimental data; the heavy central line is the fit. The thin upper
and lower lines are numerically estimated upper and lower ranges on the
theoretical fit arising from propagation of the error bars in the noise
matrix through the theory (see Materials and Methods).   Values of
($\chi^2$, $b$) for cells in (A)--(D) are (1.6, 1.1), (5.1, 0.6), (7.2,
1.0), and (1.6, 2.6), respectively.}
\label{fits}
\end{figure}

%\newpage
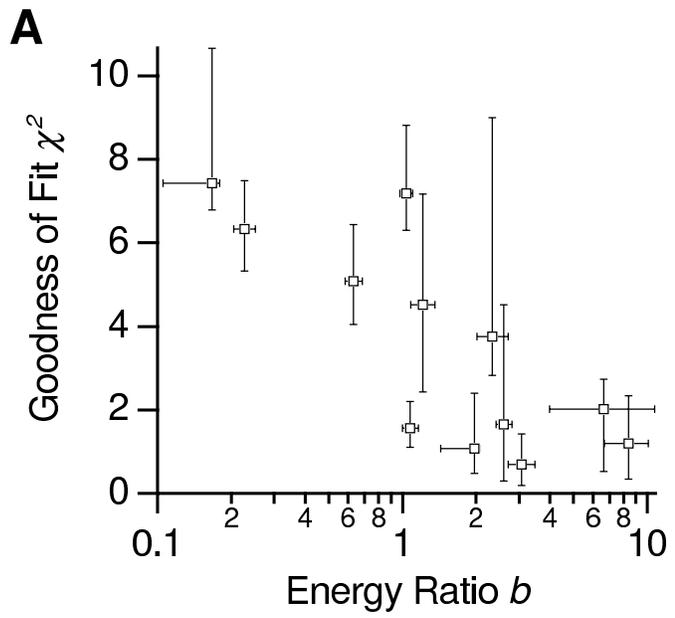
\begin{figure}[h]
%\vskip 0.5in
%\begin{center}
%\leavevmode
%\epsfxsize=6in
%\epsfysize=7.0in
%\end{center}
\caption{ (A) Goodness-of-fit ($\chi^2$) plotted versus best energy slope
($b$) for each ganglion cell.  Range bars are obtained from numerical
propagation of uncertainty in the noise matrix (see text for discussion). 
(B) Overall firing rate plotted versus energy slope for each ganglion
cell.}
\label{fitparams}
\end{figure}

%\newpage
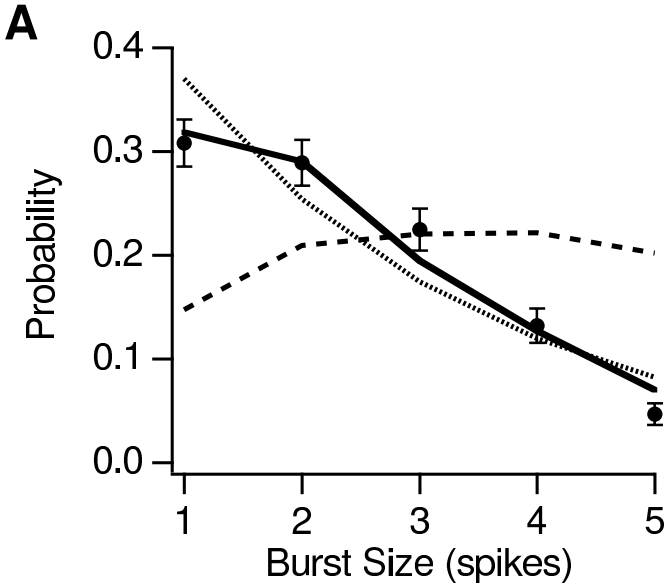
\begin{figure}[h]
%\vskip 0.5in
%\begin{center}
%\leavevmode
%\epsfxsize=6in
%\epsfysize=7.25in
%\epsffile{Fig5bphoto.eps}
%\end{center}
\caption{(A) Experimental burst size distribution (circles) for a
ganglion cell shown against the full optimal model (dark line) as well as
models without noise (dotted line) and without energy (dashed line).  The
full model is seen to match the data significantly better than either
partial model. (B) The $\chi^2$ deviation between the burst size
distribution of model and data for partial models without noise (dark
circles) and without energy (open circles) plotted against $\chi^2$ for
the full model.  The dashed line shows the case where the partial and
full models have the same error.}
\label{fitcompare}
\end{figure}

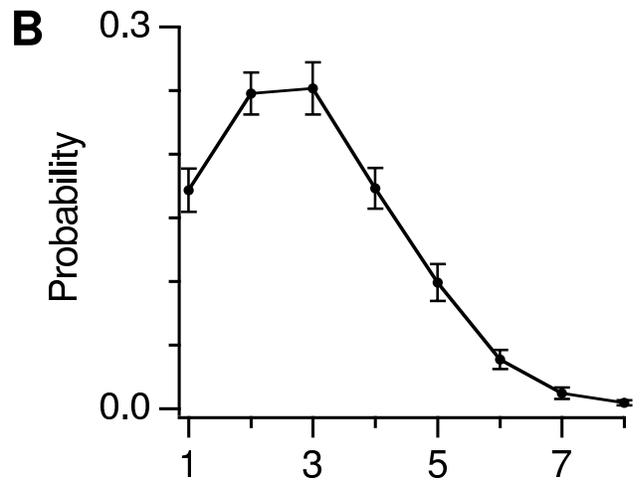
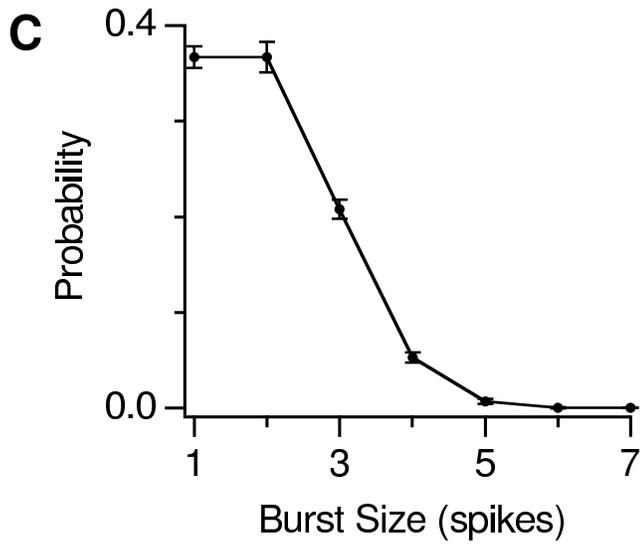
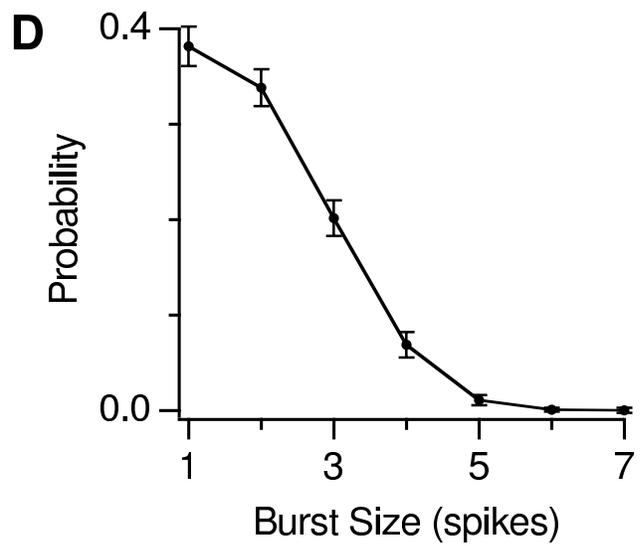
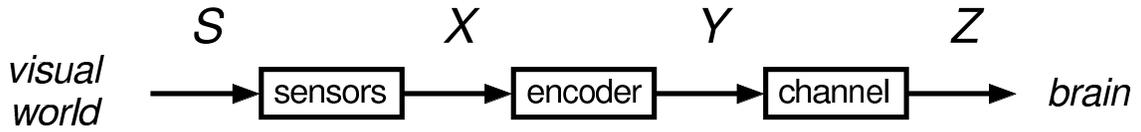
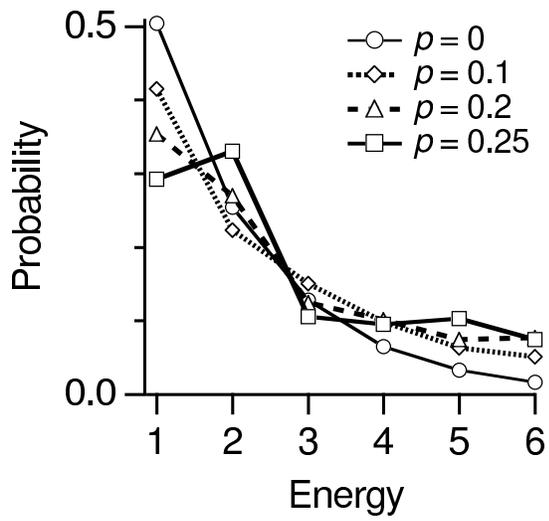
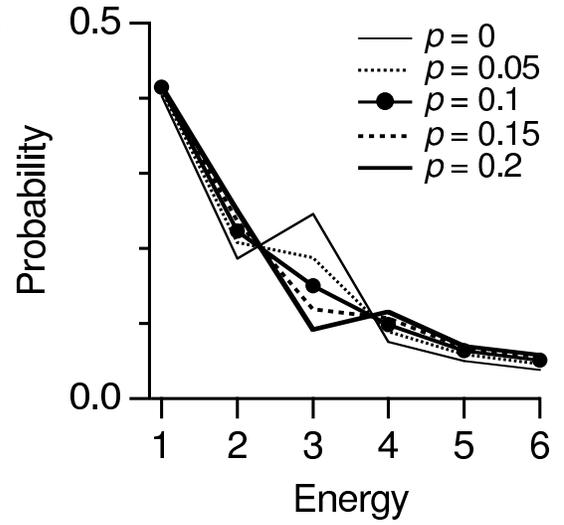
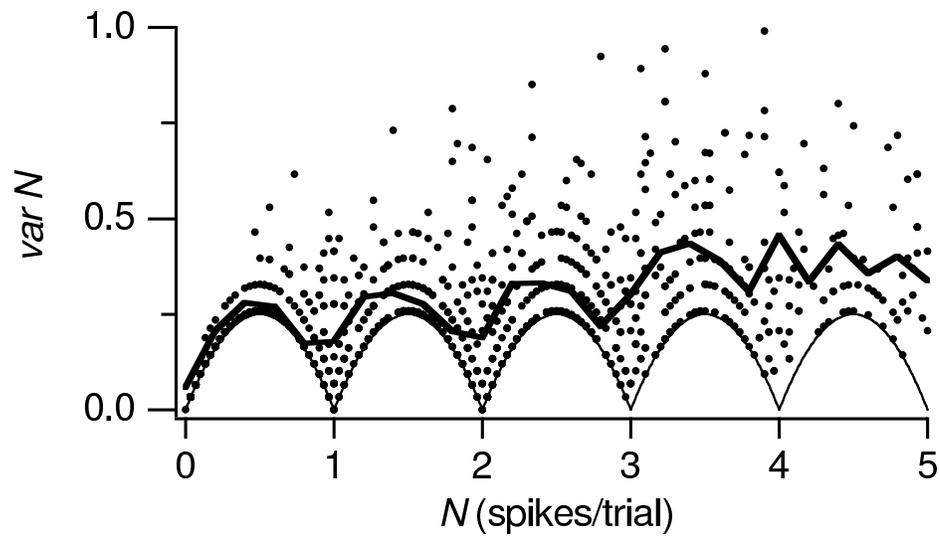
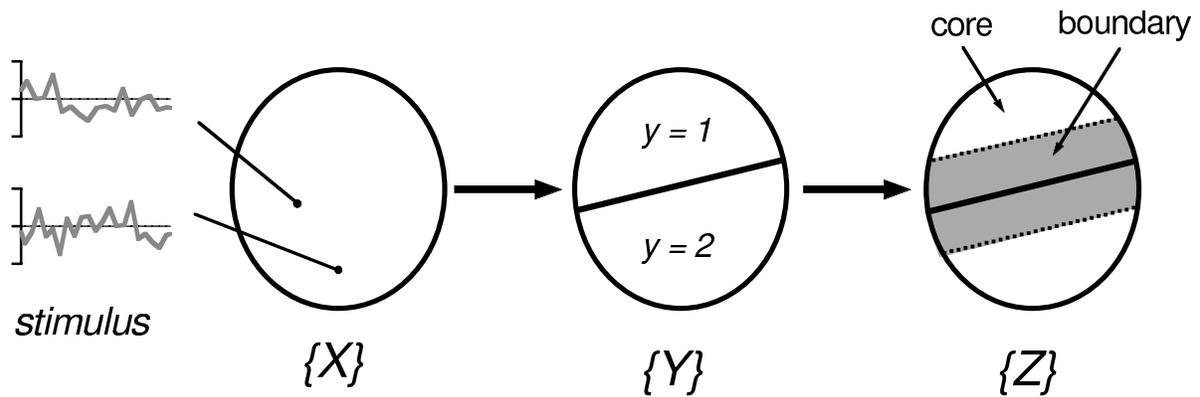
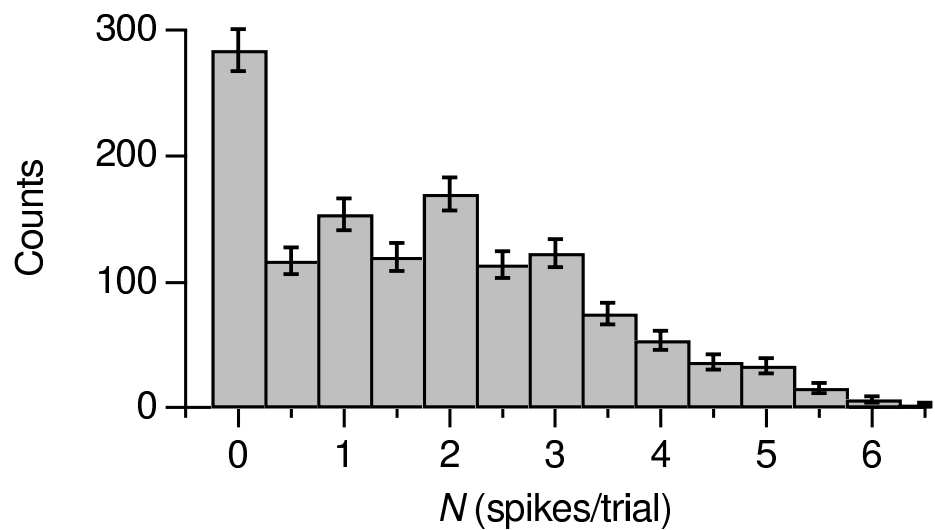
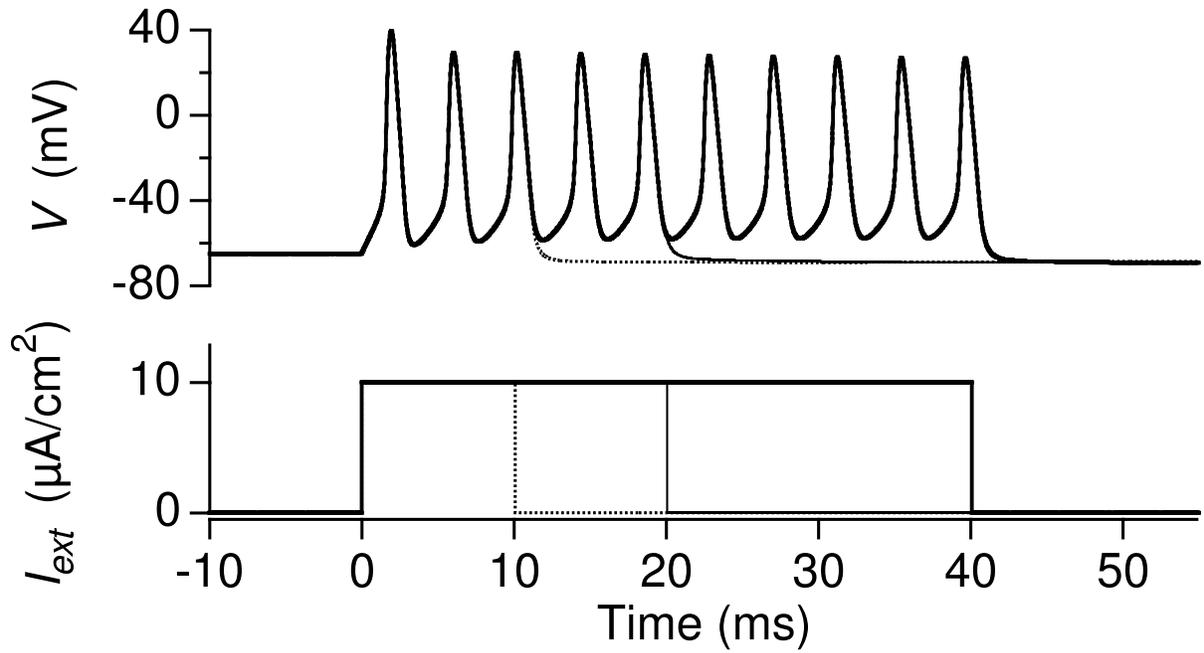
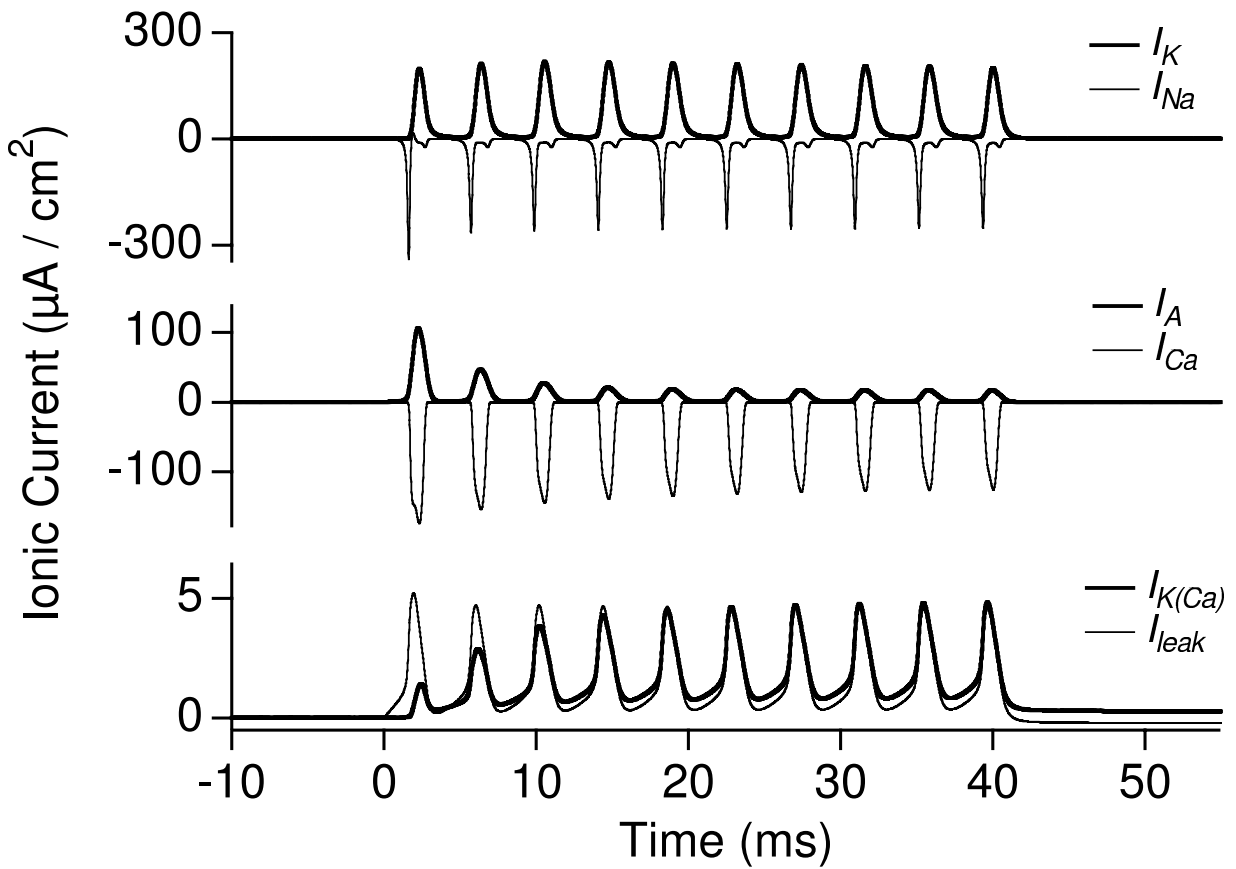
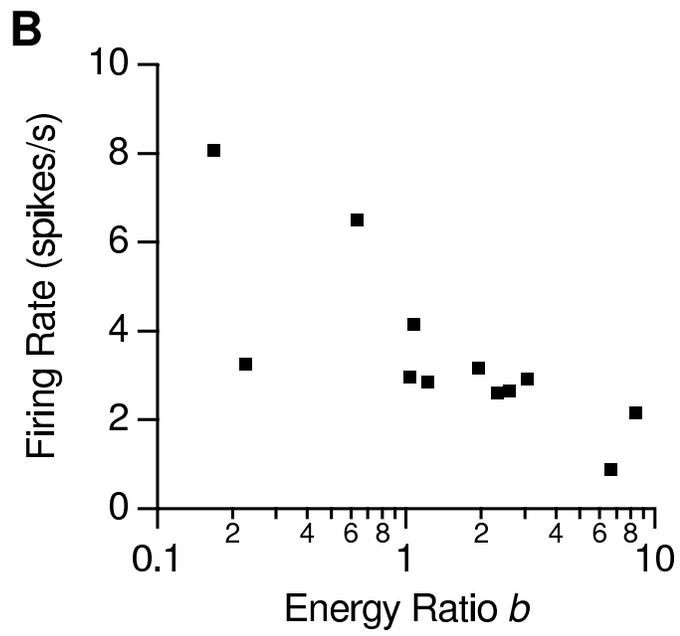
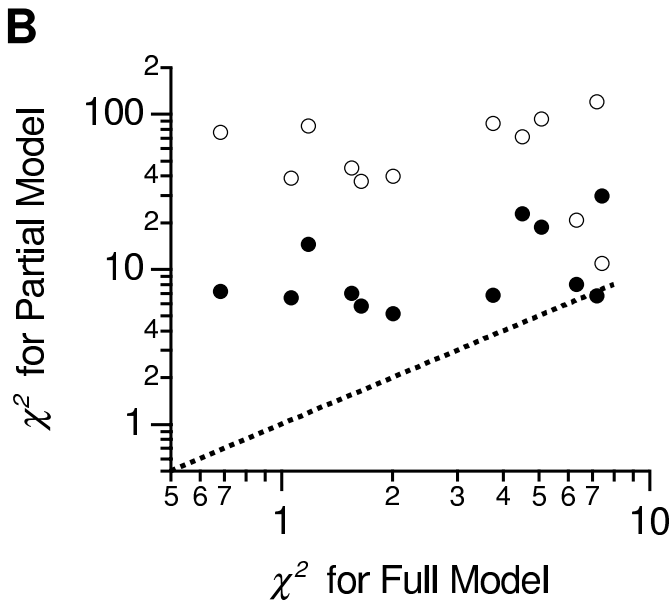
\end{document}